\documentclass[aps,preprint,groupedaddress,superscriptaddress,showpacs]{revtex4-1}

\usepackage{graphicx}
\usepackage{color}
\usepackage{amsmath}
\usepackage{amsfonts}
\usepackage{amssymb}
\usepackage{dcolumn}
\usepackage{hyperref}
\usepackage{enumerate}
\usepackage{cancel}
\usepackage{mathtools}
\usepackage{multirow} 
\usepackage{makecell}
\begin{document}

\title{Spatio-temporal characteristics of dengue outbreaks}

\author{Saulo D. S. Reis}
\email{saulo@fisica.ufc.br}
\affiliation{Departamento de F\'isica, Universidade
Federal do Cear\'a, 60451-970 Fortaleza, Cear\'a, Brazil}
\author{Lucas B\"ottcher}
\email{lucasb@ethz.ch}
\affiliation{Computational Medicine, UCLA, 90095-1766, Los Angeles, United States}
\affiliation{Institute for Theoretical Physics, ETH Zurich, 8093, Zurich, Switzerland} 
\affiliation{Center of Economic Research, ETH Zurich, 8092 Zurich, Switzerland}
\author{Jo\~{ao} P. da C. Nogueira}
\affiliation{Departamento de F\'isica, Universidade
Federal do Cear\'a, 60451-970 Fortaleza, Cear\'a, Brazil}
\author{Geziel S. Sousa}
\affiliation{Secretaria Municipal de Sa\'{u}de de Fortaleza (SMS-Fortaleza), Fortaleza, Cear\'{a}, Brazil}
\affiliation{Departamento de Sa\'{u}de Coletiva, Universidade Federal do Cear\'{a}, Fortaleza, Cear\'{a}, Brazil} 
\author{Antonio~S.~Lima~Neto}
\affiliation{Secretaria Municipal de Sa\'{u}de de Fortaleza (SMS-Fortaleza), Fortaleza, Cear\'{a}, Brazil}
\affiliation{Universidade de Fortaleza (UNIFOR), Fortaleza, Cear\'{a}, Brazil}  
\author{Hans J. Herrmann}
\affiliation{Departamento de F\'isica, Universidade
Federal do Cear\'a, 60451-970 Fortaleza, Cear\'a, Brazil}
\affiliation{PMMH. ESPCI, 7 quai St. Bernard, 75005 Paris, France}  
\author{Jos\'e S. Andrade Jr.}
\affiliation{Departamento de F\'isica, Universidade
Federal do Cear\'a, 60451-970 Fortaleza, Cear\'a, Brazil}
\date{\today}
\begin{abstract}
After their re-emergence in the last decades, dengue fever and other vector-borne diseases are a potential threat to the lives of millions of people. Based on a data set of dengue cases in the Brazilian city of Fortaleza, collected from 2011 to 2016, we study the spatio-temporal characteristics of dengue outbreaks to characterize epidemic and non-epidemic years. First, we identify regions that show a high prevalence of dengue cases and mosquito larvae in different years and also analyze their corresponding correlations. Our results show that the characteristic correlation length of the epidemic is of the order of the system size, suggesting that factors such as citizen mobility may play a major role as a drive for spatial spreading of vector-borne diseases.  Inspired by this observation, we perform a mean-field estimation of the basic reproduction number and find that our estimated values agree well with the values reported for other regions, pointing towards similar underlying spreading mechanisms. These findings provide insights into the spreading characteristics of dengue in densely populated areas and should be of relevance for the design of improved disease containment strategies.
\end{abstract}
\maketitle
\section*{Introduction}
%\noindent
%{\bf\large Introduction}\\

\begin{figure}[h!]
\begin{minipage}{\textwidth}
\centering
\includegraphics[width=\textwidth]{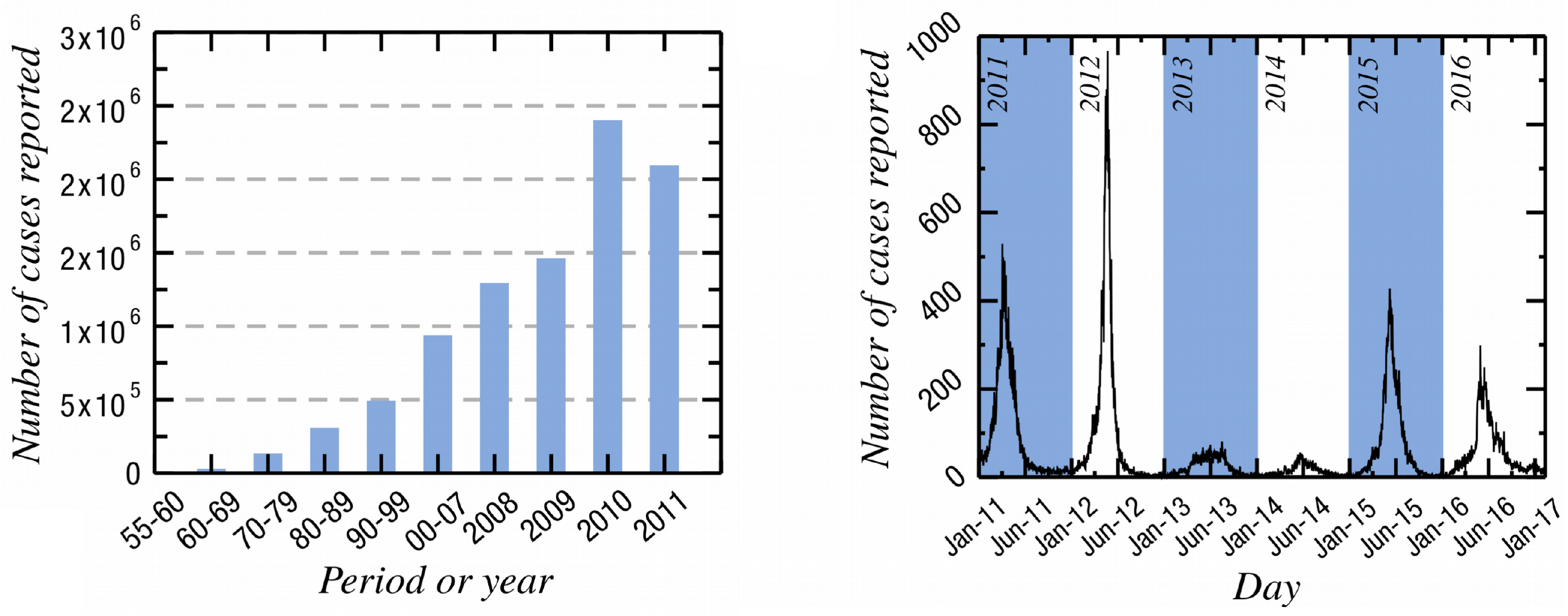}
\end{minipage}
\caption{\textbf{Temporal evolution of the number of dengue cases.} In the left panels, we show the number of dengue cases that were reported to the WHO during the last half-century~\cite{who2013,who2014}.  Between 1995--2007, the shown numbers are averages over the indicated periods. The number of cases has been increasing drastically during the last half-century. In the right panel, we show the number of reported dengue cases per day in Fortaleza from 2011 to 2016.} 
\label{fig:dengue_cases}
\end{figure}

%\noindent
According to a recent report of the world health organization (WHO), 
over 40\% of the world's population are at the risk of a dengue 
infection~\cite{who2013}. Alarmingly, in the last half-century, a 20 
to 30-fold increase of dengue cases has been monitored world-wide as shown in Fig.~\ref{fig:dengue_cases} (left). Dengue fever is a vector-borne 
disease and thus requires a disease vector to transmit the virus from 
one infected human to another susceptible one. Dengue virus 
transmission occurs through bloodsucking female \emph{Aedes aegypti} 
mosquitoes. Infected humans transmit the virus, up to maximally 12 
days after first symptoms occur, to a mosquito which further 
transmits the virus after an incubation period of 4-10
days~\cite{who2013}. Symptoms include high fever, headache, vomiting, skin rash, and muscle and joint pains~\cite{who2009}. Aedes 
aegypti is also responsible for the transmission of other severe 
vector-borne diseases such as yellow fever, chikungunya and zika 
whose recent outbreaks are challenging health officials in different 
countries~\cite{yellow_fever,chikungunya,zika_brazil}. Vector-borne 
diseases are responsible for the death of more than one million humans each year, disrupt health 
systems, and obstruct the development of many
countries~\cite{who2013}. As a consequence of the unavailability of vaccines and/or 
drug resistance, as is the case for many vector-borne diseases, 
control measures such as personal protection, reduction of vector 
breading habitats, and usage of insecticides are essential to contain 
outbreaks~\cite{who2013,neglected_disease_review}. 

Recently, studies have demonstrated the poor correspondence between levels of infestation measured by entomological surveys at national and local level~\cite{enslen2020infestation,maccormack2020evaluation,sanchez2006aedes}. The correlation between high infestation, measured by larval entomological indicators, and risk of dengue epidemic has been shown to be weak in the most diverse scenarios~\cite{cromwell2017relationship,focks1997pupal,focks2004review,gubler1998dengue,bowman2014assessing}. Despite the evidence of the low usefulness and incipient positive predictive value of these larval surveys being extensively demonstrated, some countries still consider the thresholds of infestation indices as a main indicative of epidemic risk and trigger of spatially-oriented vector control actions~\cite{enslen2020infestation,maccormack2020evaluation,bowman2014assessing}.

In fact, there is pessimism about the ability of larval indicators to be good parameters for the prediction of epidemics. Several authors have suggested the systematic use of pupal indices and those that directly measure the density of the adult mosquito, even if this implies a radical reformulation in the routines of vector control programs. They argue that the larval indexes are no longer able to fulfill their main objective, which is, when estimated the abundance of the adult vector, to anticipate the transmission of the disease~\cite{enslen2020infestation,sanchez2006aedes,focks2004review,bowman2014assessing,maccormack2018epidemiological,tun1996critical}. Moreover, pupal surveys and mosquito capture are of low cost-effectiveness and inaccurate in low infestation scenarios where the mosquito has high survival due to optimal climatic conditions and other factors, increasing vector competence and causing sustained transmission of the virus with low pupal and mosquito densities.

Urban mobility, as expressed by the movement of viremic humans, has become a more important factor than levels of infestation and abundance of the Aedes aegypti vector~\cite{bouzid2016public,stoddard2013house}. Several studies point out the importance of the human movement in the spatio-temporal dynamics of urban arboviruses transmitted by Aedes aegypti, in particular, dengue fever~\cite{vazquez2013using, vazquez2010quantifying,vazquez2009usefulness}. In 2019, the Pan American Health Organization (PAHO) published a technical document that points out that the risk stratification of transmission, using the most diverse information available, is the path to better efficiency of vector control~\cite{pan2019technical}. Theoretically, mapping cities in micro-areas with different levels of transmission risk could make vector control strategies more effective, especially in regions with limited human and financial resources~\cite{world2012handbook,quintero2014ecological,caprara2009irregular}. However, human mobility patterns are still not subject to accurate monitoring and can be crucial for the definition of high-risk spatial units that should be prioritized by the national and local programs for the control of urban arboviruses transmitted by Aedes aegypti~\cite{enslen2020infestation}.

Predicting and containing dengue outbreaks is a demanding task due to the complex nature of the underlying spreading process~\cite{guzzetta2018quantifying,antonio2017spatial}. 
Many factors such as environmental conditions or the movements of humans within certain regions influence the spreading of dengue~\cite{stolerman16,human_mobility_pnas2013,epidemics2014,human_movement_interface2017}.
The increasing danger of dengue infections makes it, however, necessary to foster a better understanding of the spreading dynamics of dengue.
In this work, we study the spatio-temporal characteristics of dengue outbreaks in Fortaleza from 2011-2016 to characterize epidemic and non-epidemic years. 
Fortaleza, the capital of Cear\'a state is one of the largest cities in Brazil and is located in the north-east of the country where dengue and other neglected tropical diseases show a high prevalence~\cite{neglected_disease_brazil,maccormack2018epidemiological}. We show in Fig.~\ref{fig:dengue_cases} (right) that up to 1000 dengue cases have been reported per day in Fortaleza during 2011 and 2016. In our analysis, we identify regions that exhibit a large number of dengue infections and mosquito larvae in different years, and also analyze the corresponding correlations throughout all neighborhoods of Fortaleza. We observe that the correlation length (i.e., the characteristic length scale of correlations between case numbers at different locations in the system) is of the order of the system size.
This provides strong support to the fact that presence factors such as citizen mobility driving the spatial spreading of the disease. 

Motivated by the observation that people interact across long distances, we use a mean-field model to compare the outbreak dynamics of two characteristic epidemic years with corresponding analyses made for other regions~\cite{pandey2013comparing}. In particular, we perform a Bayesian Markov chain Monte Carlo parameter estimation for a mean-field \emph{susceptible-infected-recovered} (SIR) model which has been previously successfully applied to the modeling of dengue outbreaks~\cite{pandey2013comparing}. Our results provide insights into the spatio-temporal characteristics of dengue outbreaks in densely populated areas and should therefore be relevant for making dengue containment strategies more effective.
\section*{Data set}
%\\\\\noindent
%{\bf \large Data set}\\
%
%
%
\begin{table}
\centering
\begin{tabular}{|c|c|}
\hline
{\bf Year} & {\bf Number of cases reported}\\ \hline
%\multicolumn{1}{|c|}{\multirow{2}{*}{\qquad \bf{Year} \qquad}} & \multicolumn{2}{c|}{\qquad \bf{Number of cases reported} \qquad\qquad}                  \\ \cline{2-3} 
%\multicolumn{1}{|c|}{}                      & \multicolumn{1}{c|}{\qquad \emph{original}\qquad\qquad} & \multicolumn{1}{c|}{\emph{processed}} \\ \hline
2011                                        & 33953\\%                         & 33836                          \\ 
2012                                        & 38319\\%                         & 38197                          \\ 
2013                                        & 8761\\%                          & 8706                           \\ 
2014                                        & 5092\\%                          & 5029                           \\ 
2015                                        & 26425\\%                         & 26176                          \\
2016                                        & 21736\\ \hline%                  & 21596                          \\ \hline
\end{tabular}
\caption{\textbf{Number of dengue cases reported in Fortaleza between 2011 and 2016.}}% The \emph{processed} subcolumn contains the \emph{original} number of cases reported after removal of cases of infected tourists and people who live in the Fortaleza metropolitan area.}
\label{table:numberofcasesreported}
\end{table}
In total, we consider three data sets that are available as \emph{Supplemental Material}. The first data set consists of spatio-temporal information on dengue infections in the city of Fortaleza from 2011 to 2016. These data have been provided by the \emph{Epidemiological Surveillance Division of Fortaleza Health Secretariat}, and contains both the date when an infected person reports a potential dengue infection to a physician and the corresponding geographic location. Serology and clinical diagnosis were used to diagnose dengue infections.
Dataset that was provided by Fortaleza Health Secretariat has been anonymized. Epidemiological and clinical variables have also been removed.
%The cases that were not diagnosed as infected with dengue have been removed from the data set. 
We %ŝalso identify and 
remove cases of infected tourists and people who live in the metropolitan area of Fortaleza. The total number of dengue cases between 2011 and 2016 in our data set after data processing is 133540. Table \ref{table:numberofcasesreported} shows the number of cases reported for each year. %before and after our first data processing. 
Our second data set contains information about Aedes aegypti larvae measurements in Fortaleza from 2011 to 2016. Measurement data is available in fortnight intervals (i.e., every two weeks) from 113 strategic points (SP) (e.g. junk yards) that are monitored by health authorities of Fortaleza over the whole city. A SP is said to be positive if Aedes aegypti larvae have been found independent of the actual amount. Data on Aedes aegypti infestation according to each Strategic Point are available on the Fortaleza Daily Disease Monitoring System~\cite{SIMDA} and were tabulated and consolidated by the epidemiological surveillance division team. Due to the influence of precipitation on the mosquito population size~\cite{stolerman16}, we also consider data from three different rain gauges in Fortaleza during 2011 to 2016.
%
%
%
%\section*{Dengue outbreaks in Fortaleza}
\\\\\noindent{\bf Dengue outbreaks in Fortaleza}\\\\
\begin{figure}
\begin{minipage}{0.49\textwidth}
\centering
\includegraphics[width=\textwidth]{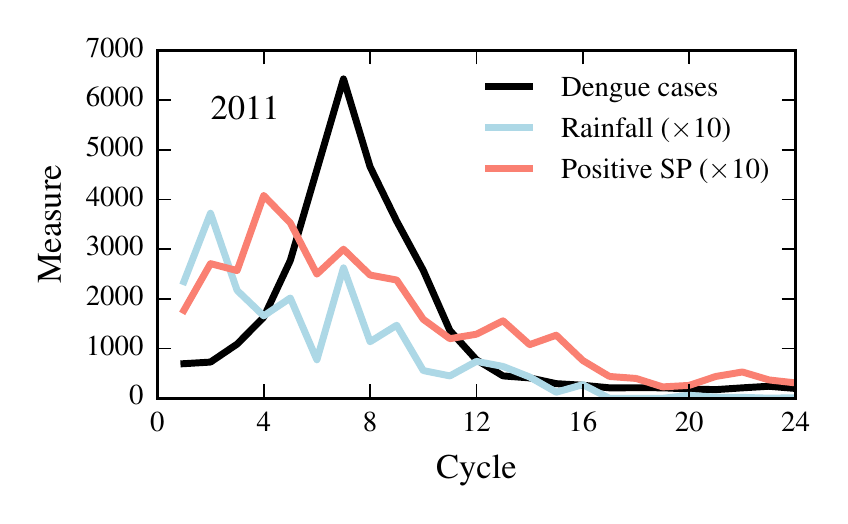}
\end{minipage}
\begin{minipage}{0.49\textwidth}
\centering
\includegraphics[width=\textwidth]{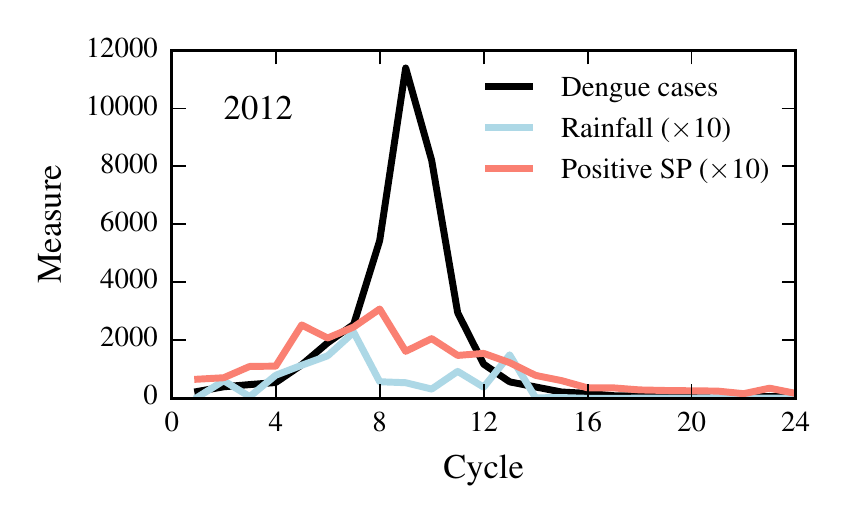}
\end{minipage}
\begin{minipage}{0.49\textwidth}
\centering
\includegraphics[width=\textwidth]{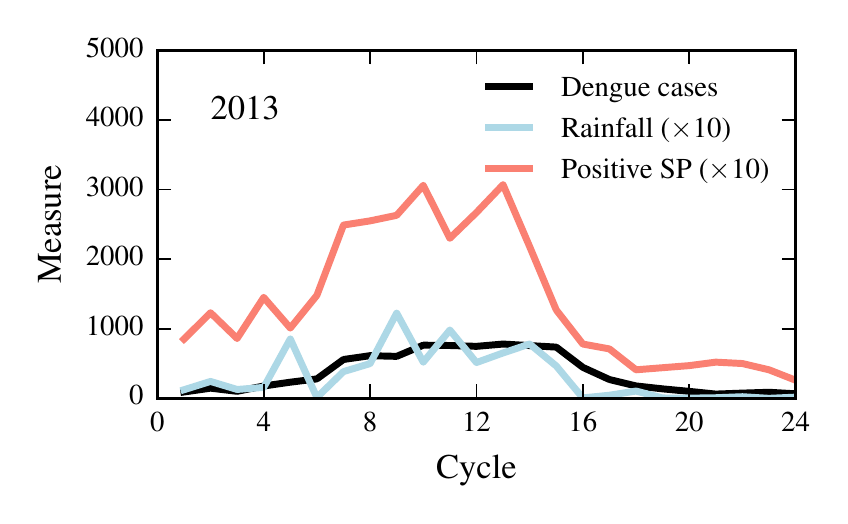}
\end{minipage}
\begin{minipage}{0.49\textwidth}
\centering
\includegraphics[width=\textwidth]{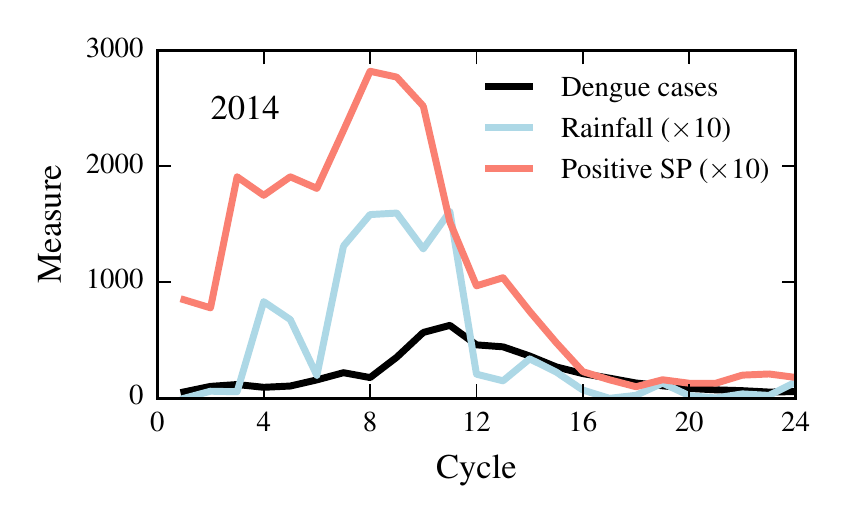}
\end{minipage}
\begin{minipage}{0.49\textwidth}
\centering
\includegraphics[width=\textwidth]{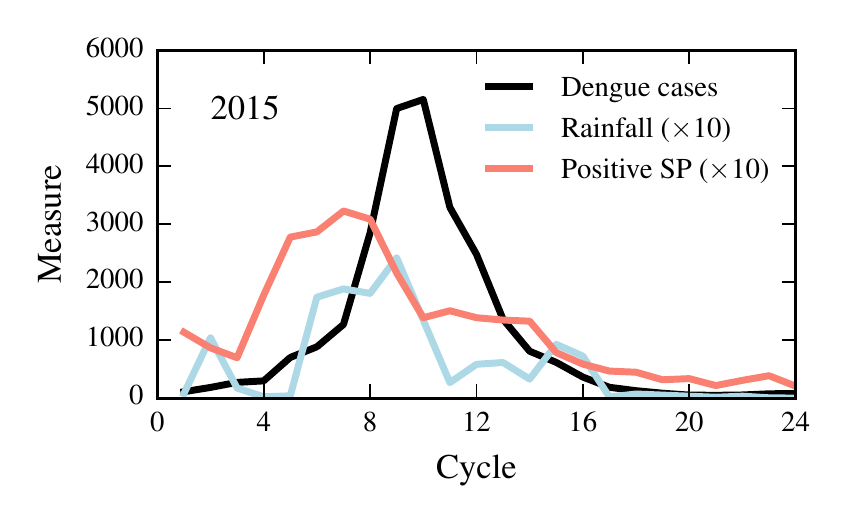}
\end{minipage}
\begin{minipage}{0.49\textwidth}
\centering
\includegraphics[width=\textwidth]{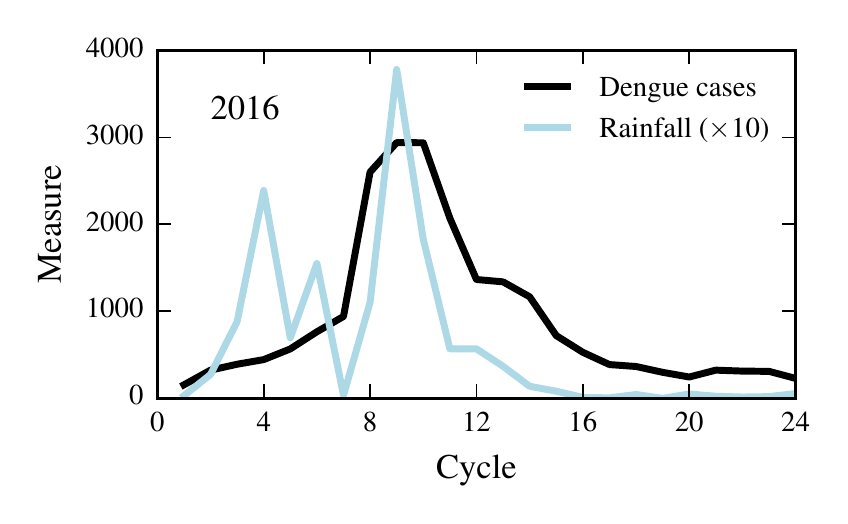}
\end{minipage}
\caption{\textbf{The number of reported dengue cases, positive SPs and rainfall as a
function of time from 2011 to 2016.}
For each year from 2011 until 2016, we show the number 
of reported dengue cases, positive SPs and the rainfall as a function of time. One time 
interval (cycle) is one fortnight. The number of positive SPs and amount of rain in liters 
per square meter have been rescaled by a factor of 10. There is no larvae measurement data 
available for 2016.} 
\label{fig:time_evolutions}
\end{figure}
\begin{table}[]
\centering
\begin{tabular}{|c|c|c|c|c|c|c|}
\hline
\multicolumn{1}{|c|}{\multirow{2}{*}{\qquad \bf{Correlations} \qquad}}

& \multicolumn{6}{c|}{\qquad \bf{Year} \qquad\qquad}                  \\ \cline{2-7} 
\multicolumn{1}{|c|}{}                      & \multicolumn{1}{c|}{\quad 2011 \quad\quad} & \multicolumn{1}{c|}{\quad 2012 \quad\quad} & \multicolumn{1}{c|}{\quad 2013 \quad\quad} & \multicolumn{1}{c|}{\quad 2014 \quad\quad} & \multicolumn{1}{c|}{\quad 2015 \quad\quad} & \multicolumn{1}{c|}{\quad 2016 \quad\quad} \\ \hline

$\tau_{\text{max}}\left(\mathrm{D}, \mathrm{R}\right)$ & 5 & 3 & 2 & 2 & 2& 0 \\ \hline
$\tau_{\text{max}}\left(\mathrm{D}, \mathrm{SP}\right)$ & 2 & 2 & 1 & 3 & 3 & -\\ \hline
$\tau_{\text{max}}\left(\mathrm{SP}, \mathrm{R}\right)$ & 2 & 1 & -1 & -1 & -1 & - \\ \hline
\end{tabular}
\caption{\textbf{The maximum correlation coefficient time lags from 2011 until 2016.} 
For each year from 2011 until 2016, we compute the correlation coefficients of the number of reported dengue cases (D), positive SPs and the rainfall (R) for different time lags $\tau$. The value of $\tau=1$ corresponds to one fortnight. We show the time lags that correspond to a maximum correlation coefficient. There is no SP data available for 2016.}
\label{table:dengue_precip_larvae_correlations}
\end{table}
According to the Brazilian ministry of health and the health administration of the prefecture of Fortaleza, the population of Fortaleza had been at a high risk of a dengue infection during the years 2011, 2012, 2015, and 2016 that are classified as epidemic years~\cite{dengue_saude}. The number of reported dengue cases in this time period is given in Table~\ref{table:numberofcasesreported} and shown in Fig.~\ref{fig:dengue_cases} (right). Considering the population size of 2.5 million~\cite{populationfortaleza}, an alarmingly high number of several hundred up to almost 1000 new infections per day have been reported during this time period. 
%The corresponding total numbers of reported dengue cases in the epidemic years are 33836, 38197, 26176, and 21596, respectively. In the non-epidemic years 2013 and 2014, the total numbers are 8706 and 5029. 
As shown in Fig.~\ref{fig:time_evolutions}, the number of reported dengue cases starts to increase in January and February just shortly after the beginning of the rain season and typically reaches its peak before July. This shows that the corresponding climate conditions facilitate the growth of mosquito populations. In addition, we also show the time evolution of the number of positive SPs (i.e., the number of strategic measurement points where Aedes aegypti larvae have been found). The data in Fig.~\ref{fig:time_evolutions} indicates that the number of positive SPs changes according to the rainfall, because the curves behave very similar although their relative amplitudes vary substantially from year to year.

To better understand the correlations between rainfall, larvae growth, and dengue cases we compute the corresponding correlation coefficients for different time lags $\tau$. A time lag of $\tau=1$ means that the two considered curves are shifted by one fortnight. In Table~\ref{table:dengue_precip_larvae_correlations}, we show the time lags $\tau_{\text{max}}$ that correspond to the maximum correlation coefficient. In the case of dengue occurrences and rainfall, we find a mean value of $\bar{\tau}_{\text{max}}\left(D,R\right)=2.3(2)$ fortnights. This result agrees with the well with findings of other studies~\cite{stolerman16} reporting  that a maximum of dengue cases will be observed a few weeks up to a few months after the rainfall maximum. The maximum correlation between dengue cases and SPs indicates that the dengue incidence reaches a maximum after $\bar{\tau}_{\text{max}}\left(D,SP\right)=2.2(8)$ fortnights after a maximum of positive SPs has been found. This result implies that the number of positive SPs may be an appropriate early warning sign to estimate when the number of dengue cases reaches its maximum. Despite the very similar curve shapes, the results are less conclusive in the case of correlations between rainfall and positive SPs.
\begin{figure*}
\begin{minipage}{1\textwidth}
\centering
\includegraphics[width=\textwidth]{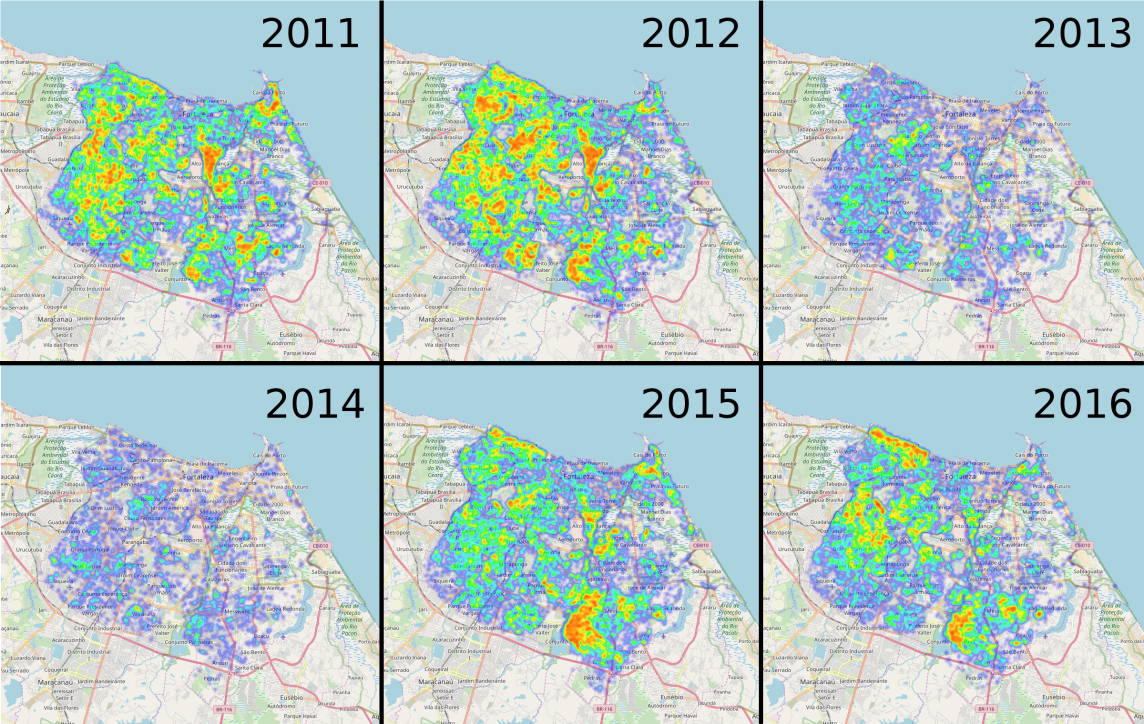}
\end{minipage}
\caption{\textbf{Dengue outbreaks in Fortaleza from 2011 until 2016.} Heat maps of all dengue cases in Fortaleza from 2011 until 2016. Blue areas correspond to regions with a low prevalence of dengue cases whereas red ones indicate large dengue outbreaks. The epidemic years are 2011, 2012, 2015 and 2016. All heat maps have been generated with Folium~\cite{folium}.} 
\label{fig:heat_map}
\end{figure*}
In some years, the maximum number of positive SPs is found after the rainfall maximum, whereas the opposite situation occurs in other years, which means that they are uncorrelated. This can be also seen by comparing the time evolutions of rainfall and the number of positive SPs in Fig.~\ref{fig:time_evolutions}.
A possible reason to explain this result may be the fact that a SP was considered as positive even if a small amount of larvae was found at this location. For this reason the positiveness of a SP indicates  the spatial extent of mosquitoes and not their concentration. Therefore, correlations between rainfall and positive SPs have to be interpreted with caution.
%%
% One reason for this result may be the fact that not the actual amounts of larvae are counted but the number of positive SPs. A SP is always counted as positive as long as at least some larvae have been found at this location. Differences between larger or smaller amounts of larvae are not considered in this measure. For these reasons the SP in fact measures the spatial extent of mosquitos and not their concentration. Therefore, correlations between rainfall and positive SPs have to be interpreted with caution. 
%
Still, the very similar shapes of the positive SP and the rainfall curves indicate that the spatial spread of the mosquito changes with the amount of rain. The influence of climate conditions on dengue outbreaks in Brazil has also been recently analyzed in reference~\cite{stolerman16}.
Due to the equatorial location of Fortaleza, the mean temperature over one year is $26.3\pm 0.6^{\rm\circ}~$C and can be regarded as constant with only limited impact on the local dengue spreading dynamics~\cite{temperaturefortaleza}.

\begin{figure*}
\begin{minipage}{1\textwidth}
\centering
\includegraphics[width=\textwidth]{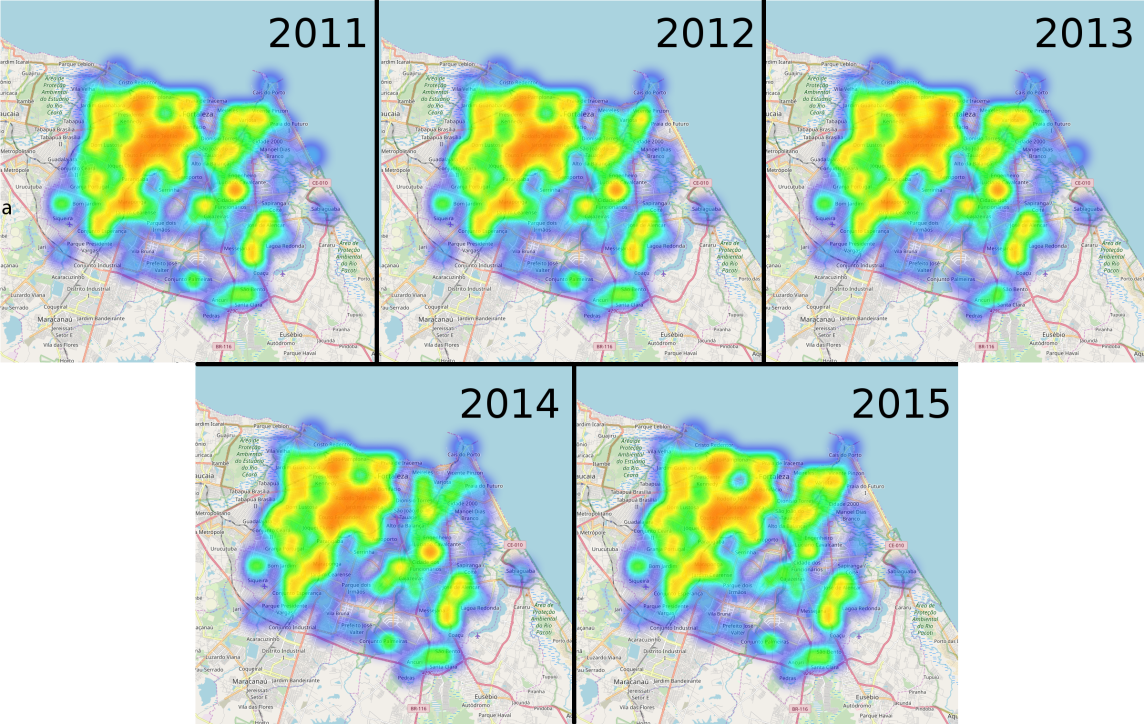}
\end{minipage}
\caption{\textbf{Positive SPs in Fortaleza from 2011 until 2015.} Heat maps of the larvae measurement outcomes in Fortaleza from 2011 until 2015. Each district has a different number of SPs. If larvae are found, the SP measurement is said to be positive. Blue areas correspond to regions with a low number of positive SPs, whereas red ones indicate the opposite. All heat maps have been generated with Folium~\cite{folium}.} 
\label{fig:heat_map_2}
\end{figure*}

In addition to the temporal information on dengue cases in Fortaleza, we collected data  about the geographical locations of all infection cases. We illustrate the geographical distribution of dengue cases in Fig.~\ref{fig:heat_map} from 2011 until 2016. If the total number of dengue infections in a certain area is large over the course of one year, this area appears in red. Areas with small numbers of dengue infections are colored blue. In 2011 and 2012 the main outbreaks occur in similar regions and cover almost the whole city except some parts in the eastern outskirts. In the non-epidemic years 2013 and 2014, the overall dengue prevalence is clearly reduced. However, some parts in the city center in the north west that also exhibit large numbers of dengue cases in the epidemic years are still at a high risk of dengue outbreaks. Also in the epidemic years 2015 and 2016, some neighborhoods in the south show a high dengue prevalence.

In addition to the geographical location of dengue cases, we also show a heat map of the spatial distribution of positive SPs in Fig.~\ref{fig:heat_map_2}. Interestingly, and in contrast to the spatial distribution of dengue cases, the majority of positive SPs are located at the same place in Fortaleza regardless of the year. In particular the city center and some parts in the north and south west of the town exhibit a large number of positive SPs. While according to Fig.~\ref{fig:heat_map_2} the regions of major outbreaks change substantially between 2011, 2013 and 2015, the number of positive SPs does not at all. Concerning the outbreak years 2015 and 2016, the recurrent dengue outbreaks and numbers of positive SPs seem to be even anti-correlated. These observations suggest that an effective disease intervention measure should target the neighborhoods which exhibit recurrent dengue outbreaks instead of numbers of positive SPs. In particular, the density of SPs is not homogeneous, which could be another reason why they fail to reproduce the major outbreaks of 2015 and 2016 in the south east of the city.  
\section*{Spatio-temporal correlations}
%\\\\\noindent{\bf Spatio-temporal correlations}\\\\
%
%
%
\begin{figure}[!ht]
\begin{minipage}{0.49\textwidth}
\centering
\includegraphics[width=\textwidth]{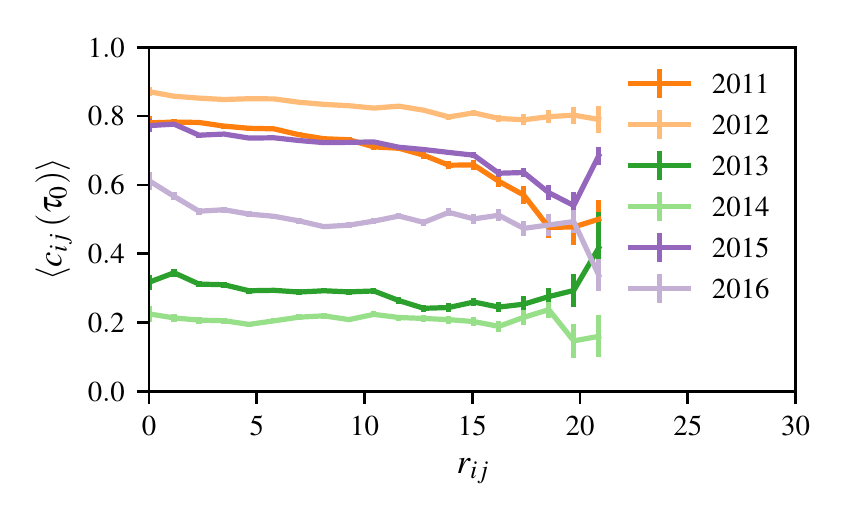}
\end{minipage}
\begin{minipage}{0.49\textwidth}
\centering
\includegraphics[width=\textwidth]{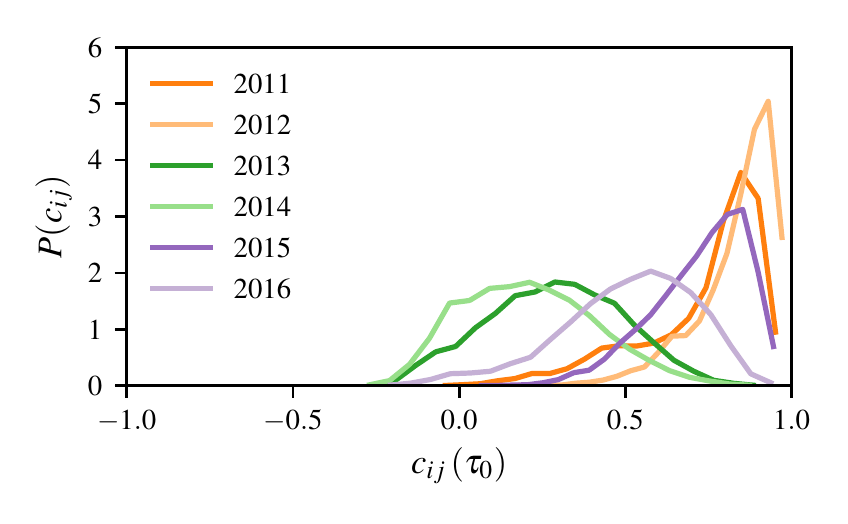}
\end{minipage}
\begin{minipage}{0.49\textwidth}
\centering
\includegraphics[width=\textwidth]{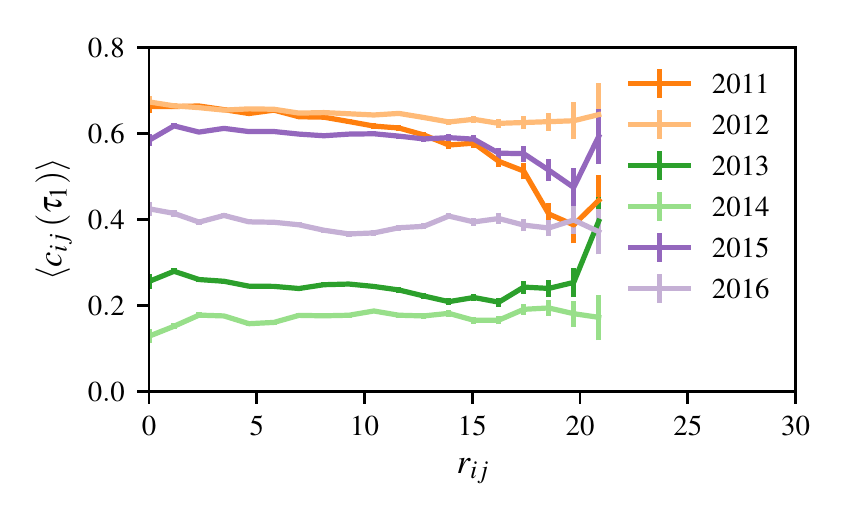}
\end{minipage}
\begin{minipage}{0.49\textwidth}
\centering
\includegraphics[width=\textwidth]{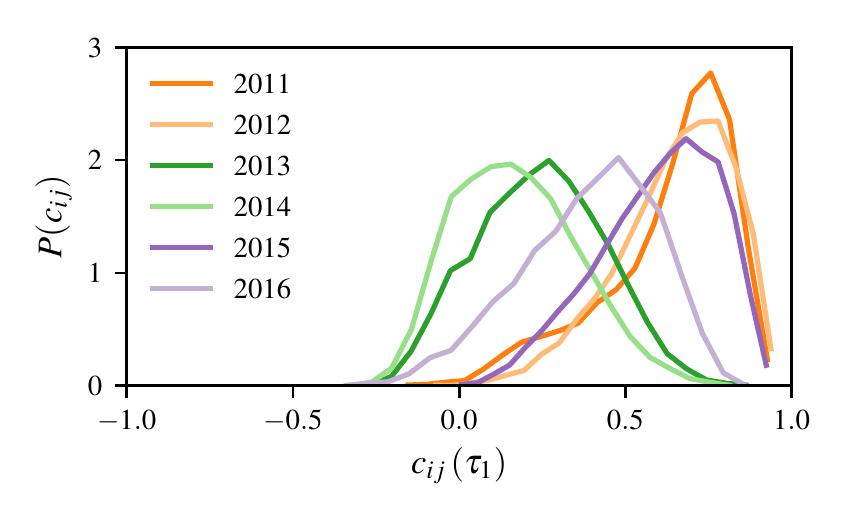}
\end{minipage}
\begin{minipage}{0.49\textwidth}
\centering
\includegraphics[width=\textwidth]{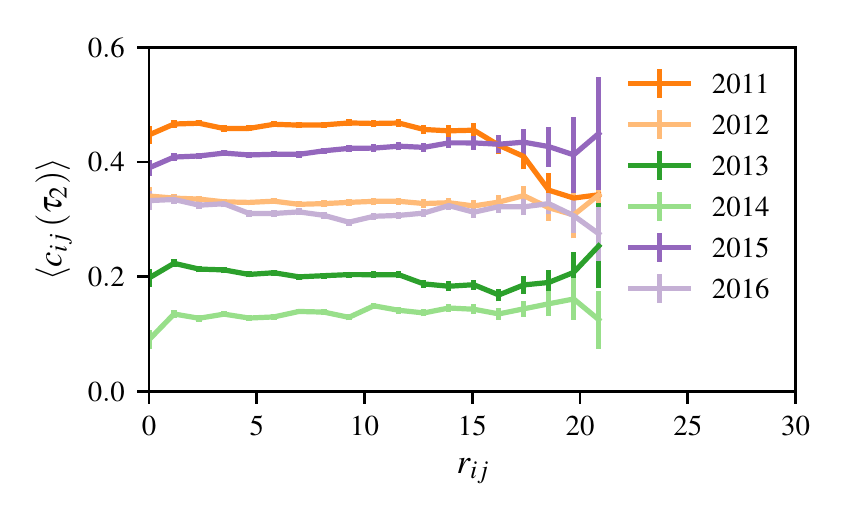}
\end{minipage}
\begin{minipage}{0.49\textwidth}
\centering
\includegraphics[width=\textwidth]{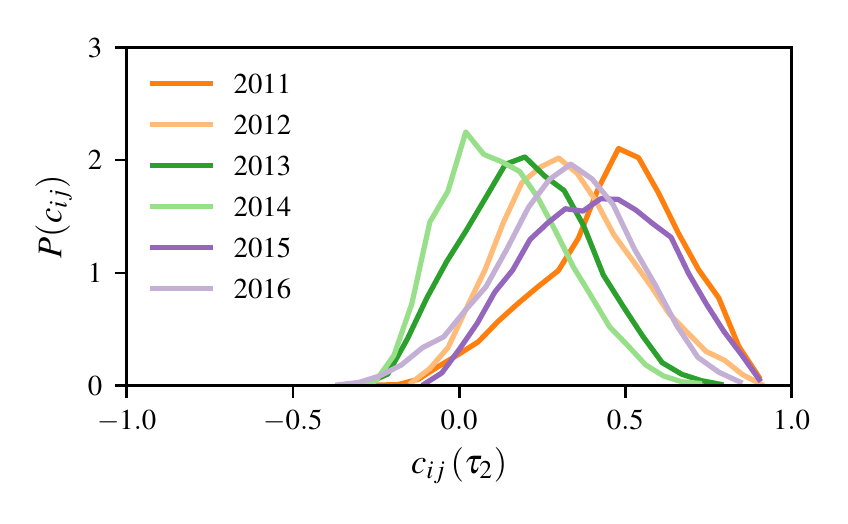}
\end{minipage}
\caption{\textbf{Correlations for different time lags, distances and years and their corresponding distributions.} The left panels show average of the correlation coefficients $\langle c_{ij}(\tau)\rangle$ as a function of the distance $r_{ij}$ between neighborhoods $i$ and $j$ time lags of $\tau_0=0$ weeks, $\tau_1=2$ weeks and $\tau_2=4 $ weeks. As depicted, epidemic years (2011, 2012, 2015, and 2016) present a higher value of $\langle c_{ij}(\tau)\rangle$ when compared to non-epidemic years (2013 and 2014). The right panels show the corresponding distribution $P\left(c_{ij}\right)$ of the correlation coefficients $c_{i j}(\tau, r_{ij})$. Epidemic years present a negative skewness for $P(c_{ij})$, while non-epidemic years have no skewness.} 
\label{fig:correlation_coefficients}
\end{figure}
For a better understanding of the spreading dynamics of dengue in urban environments, it is crucial to analyze the influence of local dengue outbreaks on outbreaks at other locations and vice-versa. In this way,
%
%We have shown the spatial distribution of cummulated dengue cases in Fortaleza in Fig.~\ref{fig:heat_map} without analyzing the influence of local dengue outbreaks on outbreaks at other locations. To better understand this possible influence, 
%
We therefore now focus on spatio-temporal correlations between dengue incidences of different districts. In total, there are 118 neighborhoods in Fortaleza, and for each neighborhood $i$ we consider the number of reported dengue cases $\mathrm{NC}_i^t$ at time $t$ and the corresponding population $\mathrm{POP}_i$. The disease prevalence in neighborhood $i$ at time $t$ is given by
\begin{equation}
p_i^t=\frac{\mathrm{NC}_i^t}{\mathrm{POP}_i}.
\label{eq:disease_prevalence}
\end{equation}
We characterize the spatio-temporal correlations of dengue outbreaks between sites $i$ and $j$ by
\begin{equation}
c_{ij}(\tau, r_{ij})=\frac{1}{T} \frac{\sum_{t=1}^T\left(p_i^{t-\tau}-\langle p_i\rangle \right)\left(p_j^{t}-\langle p_j\rangle \right)}{\sigma_1 \sigma_2},
\label{eq:correlator}
\end{equation}
where $\tau$ denotes a time lag, $r_{ij}$ is the distance between neighborhood $i$ and neighborhood $j$, and $\sigma_i^2= T^{-1} \sum_{t=1}^T \left(p_i^t-\langle p_i \rangle \right)^2$ represents the variance. In our following analysis, we consider two time intervals: (i) four weeks (i.e., $T=12$) and (ii) two weeks (i.e., $T=26$).

Dengue outbreaks that occur in a certain region of the city may lead to outbreaks in another more distant region due to human mobility, because mosquitoes at the new location then have the possibility of getting in contact with the virus~\cite{human_mobility_pnas2013,epidemics2014,human_movement_interface2017}.
To analyze this effect and identify characteristic correlation length-scales, we study the correlation between the dengue prevalence in neighborhood $i$ and another neighborhood $j$ located within a distance of $r_{i j}$. Here, the distance $r_{i j}$ is the distance between the barycenters of neighborhoods $i$ and $j$ extracted from their geographical contours. Specifically, we study the correlations between neighborhoods $i$ and $j$, i.e.~$c_{ij}(\tau, r_{ij})$ as defined by Eq.~\eqref{eq:correlator}, for different time lags $\tau$ and radii $r_{i j}$. According to the definition of $c_{i j}(\tau, r_{ij})$ in Eq.~\eqref{eq:correlator}, a correlation length $\xi$ smaller than our considered system size would lead to an observable decay of $c_{ij}(\tau, r_{ij})$ for $r_{i j} > \xi$. In all considered years, the correlations $c_{i j}(\tau)$ only vary slightly with the 
distance $r_{i j}$, as shown in Fig.~\ref{fig:correlation_coefficients}. Within the error bars of our data we do not observe a substantial decay 
of $c_{i j}(\tau)$ for values of $r_{i j}$ smaller than the system size of about $15$~km. We thus conclude that the characteristic correlation length $\xi$ is of the order of 
the system size.

We choose three time lags, namely, $\tau_0=0$ weeks, $\tau_1 = 2~\mathrm{weeks}$ and $\tau_2=4~\mathrm{weeks}$ which are of the order of the transmission time scale of 12 days~\cite{who2013}. The dependence of the correlations on different time lags, distances and years is shown in Fig.~\ref{fig:correlation_coefficients} (left). For no time lag, $\tau_0=0$~weeks, or a time lag of $\tau_1=2~\mathrm{weeks}$, we find that the correlations are substantially larger in the epidemic years 2011, 2012, 2015 and 2016 as compared to the non-epidemic years 2013 and 2014. However, in 2016 the correlations are less pronounced compared to other epidemic years since the outbreaks are widely distributed and their densities rather small, as shown in Fig.~\ref{fig:heat_map}. In the case of the larger time lag, $\tau_2=4~\mathrm{weeks}$, the average correlation $\langle c_{i j}(\tau_2)\rangle$ for epidemic years decreases when compared with smaller time lags. This behavior is in accordance with observation, since a time lag of $4~\mathrm{weeks}$ exceeds the dengue transmission period, and we expect to find smaller correlations. In Fig.~\ref{fig:correlation_coefficients} (right), we show the corresponding distributions of $P(c_{i j})$ and find that they allow to clearly distinguish between epidemic and non-epidemic years for no time lag and a time lag $\tau_1=2$~weeks. In particular in the case of no time lag, the correlations are substantially larger in epidemic years compared to the non-epidemic ones.

%Based on the definition of $c_{i j}(\tau)$ in Eq.~\eqref{eq:correlator}, 
%this implies that the disease densities $p_i^t$ and $p_j^t$ must be sufficiently large for 
%every radius $r_{i j}$ during the same time interval $t$. 
It is unlikely that disease 
vectors are responsible for such correlation effects over distances of multiple kilometers due to their limited movement capabilities. In particular, it is known that the maximum flight distance reached by the Aedes aegypti from its breeding location is of the order of approximately 100~meters. On the other hand, humans travel through the densely populated urban region on a daily basis, and may transfer the virus to Aedes aegypti mosquitoes at different locations. This virus transfer mechanism is therefore compatible with the large correlation lengths revealed in this study and also agrees well with recent studies that suggest that human mobility is a key component of dengue spreading~
\cite{human_mobility_pnas2013,epidemics2014,human_movement_interface2017}.
\section*{Estimating disease transmission parameters}
%\\\\\noindent{\bf Estimating disease transmission parameters}\\\\
%
%
%
As mentioned in the previous section, the considered dengue outbreaks in Fortaleza exhibit a correlation length $\xi$ which is of the order of the underlying system size. Based on this observation, we apply mean-field epidemic model, as a first approximation, to further characterize the observed spreading dynamics. In particular, we aim at comparing the disease transmission parameters of dengue outbreaks in Fortaleza with the results of previous studies~\cite{pandey2013comparing}. To do so, we consider the two epidemic years 2012 and 2015 and perform a Bayesian Markov chain Monte Carlo parameter estimation for a SIR model that has been previously applied in the context of dengue outbreaks in Thailand~\cite{pandey2013comparing}. More details about the methodology are presented in Appendix~\ref{sec:app_sir}. Such a parameter estimation enables us to determine the basic reproduction number $R_0$ of dengue outbreaks in Fortaleza and compare it with the values of other disease outbreaks. The basic reproduction number constitutes an important epidemiological measure and is defined as the average number of secondary cases originating from one infectious individual during the initial outbreak period (i.e., in a fully susceptible population)~\cite{keeling-rohani2008}. Different models exist to characterize and predict epidemic outbreaks~\cite{keeling-rohani2008,boettcher14,boettcher16,boettcher171,bottcher2017targeted}.

In the case of dengue, some models explicitly incorporate a mosquito population, whereas others take into account such effects by using an effective spreading rate~\cite{pandey2013comparing,fitzgibbon2017outbreak}. According to Ref.~\cite{pandey2013comparing}, an explicit treatment of vector populations just increases the number of modeling parameters and may not lead to a better agreement between model and data. We therefore do not explicitly model a vector population and consider a SIR model with an effective spreading rate which has been found to capture essential features of dengue outbreaks in Thailand~\cite{pandey2013comparing}. The governing equations are
\begin{align}
\frac{\mathrm{d} S}{\mathrm{d} t} &= \mu_H N -\beta \frac{I}{N} S-\mu_H S, \label{eq:sir_s} \\
\frac{\mathrm{d} I}{\mathrm{d} t} &= \beta \frac{I}{N} S-\gamma_H I-\mu_H I, \label{eq:sir_i} \\
\frac{\mathrm{d} R}{\mathrm{d} t} &= \gamma_H I - \mu_H R, \label{eq:sir_r}
\end{align}
where $N=S+I+R$ is the human population size and $S=S(t)$, $I=I(t)$, $R=R(t)$ the numbers of susceptible, infected and recovered individuals at time $t$, respectively. The human mortality rate is denoted by $\mu_H$ and the human recovery rate by $\gamma_H$. The term $\mu_H N$ in Eq.~\eqref{eq:sir_s} corresponds to the birth rate and is chosen such that the population size is kept constant. Furthermore, the composite human-to-human transmission rate $\beta$ accounts for transitions from susceptible to infected. The relation
\begin{equation}
\beta\approx \frac{m c^2 \beta_H \beta_V}{\mu_V}
\end{equation}
connects $\beta$ with the mosquito-to-human transmission rate per bite $\beta_H$, human-to-mosquito transmission rate $\beta_V$, number of mosquitoes per person $m$, mean rate of bites per mosquito $c$, and mosquito mortality rate $\mu_V$~\cite{pandey2013comparing}.
The basic reproduction number of the SIR model is~\cite{keeling-rohani2008}
\begin{equation}
R_0=\frac{\beta}{\mu_H+\gamma_H}.
\label{eq:R0}
\end{equation}
For $R_0>1$ there exists a stable endemic state whereas the disease-free equilibrium is stable for $R_0\leq 1$. In addition to Eqs.~\eqref{eq:sir_s}-\eqref{eq:sir_r}, we define the time evolution of the cumulative number of reported dengue cases $C=C(t)$ by
\begin{equation}
\frac{\mathrm{d} C}{\mathrm{d} t}=p \beta \frac{I}{N} S,
\label{eq:sir_c}
\end{equation}
where $p$ is the fraction of infected individuals that were diagnosed with dengue and reported to the health officials. We set the population size to $N=2.4$~million in agreement with the most recent census data of Fortaleza~\cite{populationfortaleza}. Moreover, we use a mortality rate of $\mu_H=1/76~\mathrm{y}^{-1}$ in accordance with the latest World Bank life expectancy estimates~\cite{lifeexpectancy}.

A parameter estimation based on Eqs.~\eqref{eq:sir_s}-\eqref{eq:sir_r} and \eqref{eq:sir_c} means that we have to determine the parameters $\beta$, $\gamma_H$, $p$ and the initial fraction of recovered individuals $r_0=R(0)/N$ that best describe the observed dengue outbreaks. More specifically, we use Bayes' theorem to determine the posterior parameter distribution
\begin{equation}
P\left(\theta|D\right)\propto P\left(D|\theta\right)P\left(\theta\right)
\label{eq:bayes}
\end{equation}
based on the likelihood function $P\left(D|\theta\right)$ (i.e., the conditional probability of obtaining the data $D$ for given model parameters $\theta$) and the prior parameter distribution $P\left(\theta\right)$. The actual Bayesian Markov chain Monte Carlo algorithm is described in Appendix~\ref{sec:app_sir}.

\begin{figure}
\begin{minipage}{0.49\textwidth}
\centering
\includegraphics[width=\textwidth]{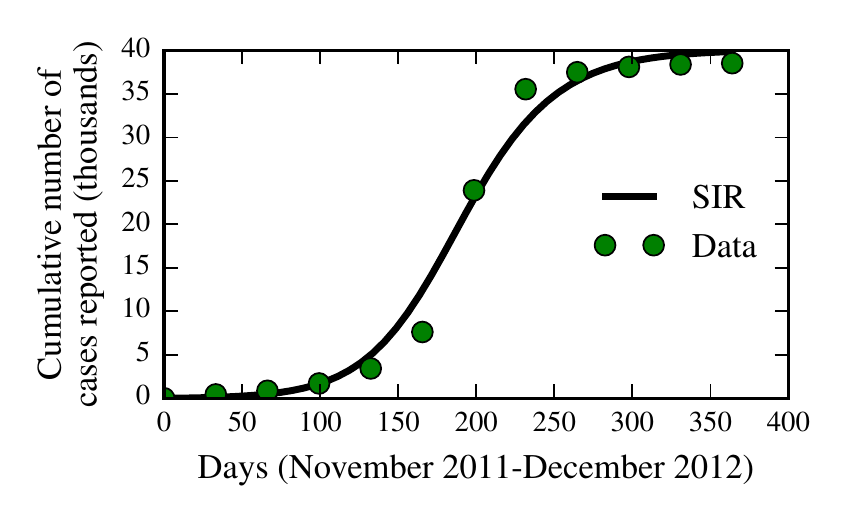}
\end{minipage}
\begin{minipage}{0.49\textwidth}
\centering
\includegraphics[width=\textwidth]{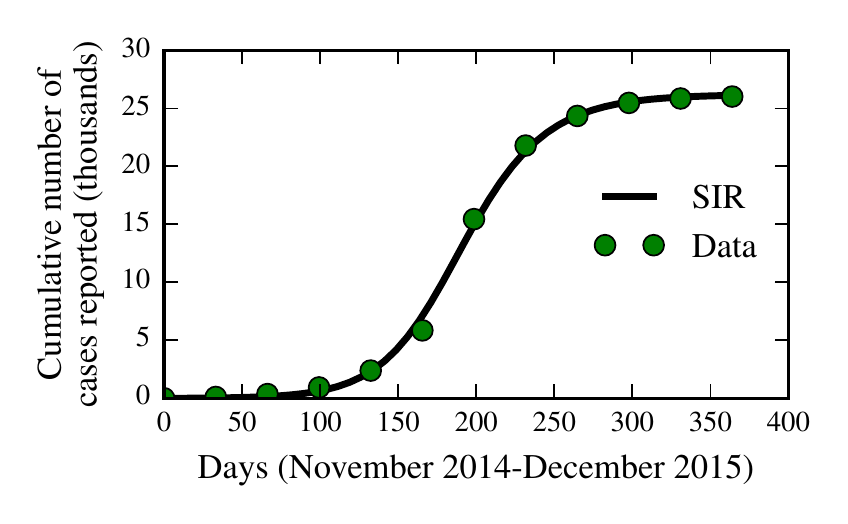}
\end{minipage}
\caption{\textbf{SIR model fit for the epidemic years 2012 and 2015.} The maximum likelihood fits of the SIR model (black solid lines) as defined by Eqs.~\eqref{eq:sir_s}-\eqref{eq:sir_r} and \eqref{eq:sir_c} for the epidemic years 2012 (left) and 2015 (right). The cumulative number of reported dengue cases according to our data set is indicated by green circles.} 
\label{fig:sir_ml_fit}
\end{figure}

In order to be able to compare our results with the ones of Ref.~\cite{pandey2013comparing} and to obtain a full infection period, we consider November as our starting month, because the rainfall then typically reaches a minimum. 
In Table~\ref{table:sirparams}, we summarize the inferred model parameters, median values, and 95\% confidence intervals of the posterior distributions. In addition, we also present the maximum likelihood (ML) estimates which we use to compare the SIR model with the actual dengue case data in Fig.~\ref{fig:sir_ml_fit}. We find good agreement between the model prediction and the actually reported number of dengue cases. Only the dengue outbreak peak between April and June 2012 is difficult to capture because due to overwhelming number of up to 1000 new infections per day.
\begin{figure}[!ht]
\begin{minipage}{0.49\textwidth}
\centering
\includegraphics[width=\textwidth]{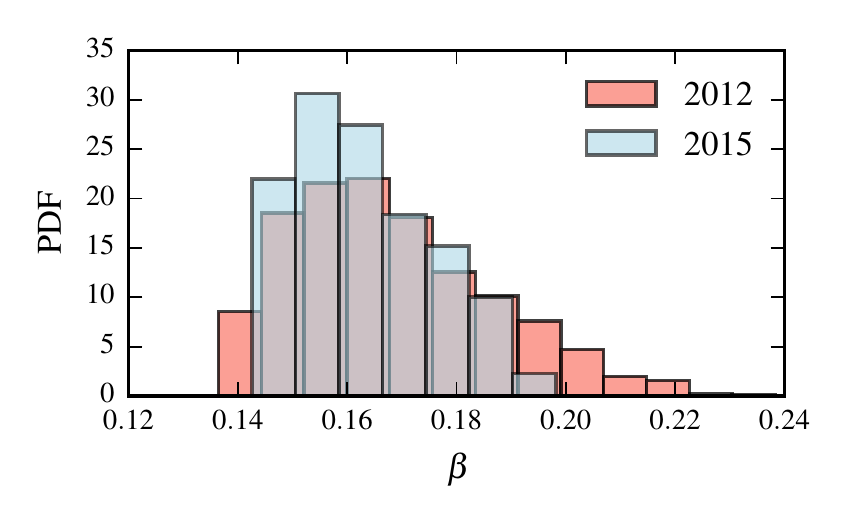}
\end{minipage}
\begin{minipage}{0.49\textwidth}
\centering
\includegraphics[width=\textwidth]{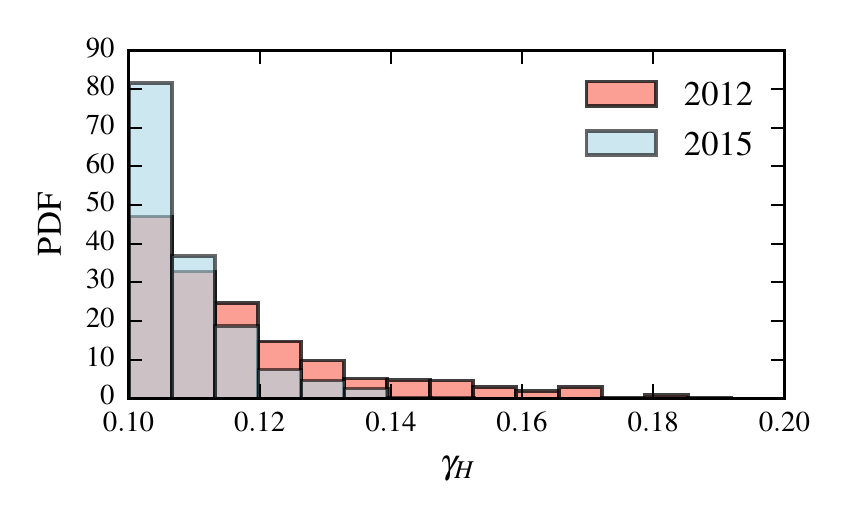}
\end{minipage}
\begin{minipage}{0.49\textwidth}
\centering
\includegraphics[width=\textwidth]{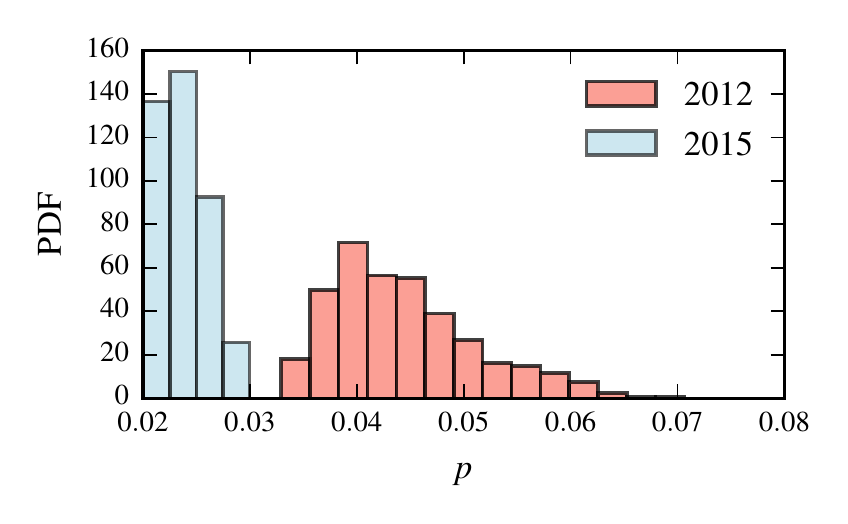}
\end{minipage}
\begin{minipage}{0.49\textwidth}
\centering
\includegraphics[width=\textwidth]{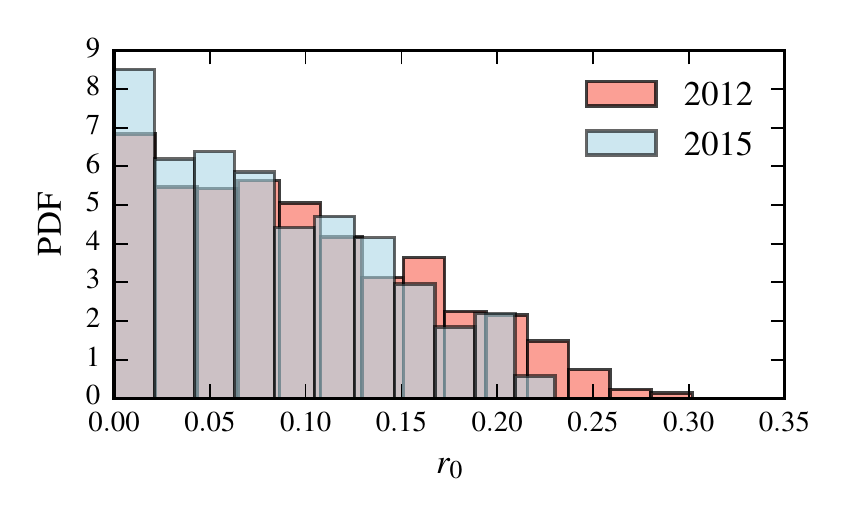}
\end{minipage}
\caption{\textbf{Posterior distributions of SIR model parameters.} The probability density functions (PDFs) of the posterior distributions of the modified SIR model as defined by Eqs.~\eqref{eq:sir_s}-\eqref{eq:sir_r} and \eqref{eq:sir_c}. The two epidemic years 2012 (orange bars) and 2015 (blue bars) have been considered for the parameter estimation.} 
\label{fig:sir_posterior}
\end{figure}

After estimating the SIR model parameters for the epidemic years 2012 and 2015, we now compare the obtained estimates with the ones of Ref.~\cite{pandey2013comparing} which focuses on dengue outbreaks in Thailand. In ref~\cite{pandey2013comparing}, the number of reported dengue hemorrhagic fever (DHF) cases, a more severe form of dengue fever, has been used, while we considered the number of all reported dengue cases. The total number of mentioned DHF cases in Thailand is roughly $70000$ with a total population size of $46.8$ million in 1984~\cite{pandey2013comparing}. In contrast to these results, the number of reported dengue cases in Fortaleza in 2012 is almost $40000$ with a total population size of $2.5$ million. This alarming difference may be partially a result of the different definitions of reported cases (DHF versus dengue fever) as well as due to sub-notification, but it also points out to the need of better control measures to contain the outbreaks in Fortaleza. Overall, the parameter estimation for the epidemic years 2012 and 2015 leads to values in a similar range compared to the results of Ref.~\cite{pandey2013comparing}. The ML estimates of $p$ and $r_0$ are a factor 2-3 larger in our parameter estimation.
On the other hand, the ML estimates of $\beta$ and $\gamma$ are slightly smaller in Fortaleza. The basic reproduction number as defined in Eq.~\eqref{eq:R0} (basically the fraction of $\beta$ and $\gamma_H$) is $1.44$ (2012) and $1.50$ (2015) and thus exhibits a value within the confidence interval of Ref.~\cite{pandey2013comparing}. This result suggests that the dengue outbreaks in both locations show similar characteristics. In other words, the average number of secondary dengue cases originating from one infection is almost identical in both locations.
\begin{table}[]
\centering
\begin{tabular}{|c|c|c|c|}
\hline
{\bf Parameter} & \quad {\bf ML} \quad & \quad {\bf Median} \quad & \quad {\bf 95\% CI} \quad \\ \hline
\makecell{$\beta$ $\left(d^{-1}\right)$ \\ Composite transmission rate} & \makecell{0.1453 (2012) \\ 0.1518 (2015)} & \makecell{0.1648 (2012) \\ 0.1606 (2015)} & \makecell{(0.1409, 0.2138) (2012) \\ (0.1453, 0.1966) (2015) } \\ \hline
\makecell{$\gamma_H$ $\left(d^{-1}\right)$ \\ Human recovery rate} &  \makecell{0.1011 (2012)\\0.1009 (2015)} & \makecell{0.1121 (2012)\\ 0.1064 (2015)} & \makecell{(0.1005, 0.1650) (2012) \\ (0.1004, 0.1299) (2015)} \\ \hline
%\makecell{$\mu_H$ $\left(y^{-1}\right)$ \\ Human mortality rate} & - & - & - \\ \hline
\makecell{$p$ \\ Probability of reporting a dengue case} & \makecell{0.0368 (2012) \\ 0.0218 (2015)} &\makecell{ 0.0432 (2012) \\ 0.0234 (2015)} & \makecell{(0.0348, 0.0609) (2012) \\ (0.0206, 0.0291) (2015)} \\ \hline
\makecell{$r_0$ \\ Initial fraction of humans recovered} & \makecell{0.0594 (2012) \\ 0.0618 (2015)} & \makecell{0.0769 (2012) \\ 0.0636 (2015)} & \makecell{(0.0066, 0.2928) (2012) \\ (0.0045, 0.2609) (2015)}\\ \hline
\makecell{$R_0$ \\ Basic reproduction number} & \makecell{1.4367 (2012) \\ 1.5039 (2015)} & \makecell{1.5384 (2012) \\ 1.4904 (2015)} & \makecell{(1.2184, 1.9584) (2012) \\ (1.3471, 1.7396) (2015)}\\ \hline
\end{tabular}
\caption{\textbf{Posterior parameter estimations.} Based on given prior SIR parameter distributions, the corresponding posterior distributions for the dengue outbreaks in 2012 and 2015 have been obtained using Bayesian Markov chain Monte Carlo sampling.
%$U\left(\cdot \right)$ is a uniform prior with the specified bounds. 
For both years an additional maximum likelihood (ML) parameter estimate is given. The posterior distributions are characterized by their median values and their 95\% confidence intervals (CI).}
\label{table:sirparams}
\end{table}
\section*{Discussion}
%\\\\\noindent{\bf\large Discussion}\\\\
%
%
%
The re-emergence of dengue and other neglected tropical diseases is a major threat in many countries. We analyzed the recent dengue outbreaks in Fortaleza as one of the largest Brazilian cities. We identified regions which exhibit a large number of dengue infections and Aedes aegypti larvae over different years. %In the future, these regions should be targeted by appropriate disease control measures. 
Our results show that the characteristic length scale of correlation effects between the number of cases at different locations is of the order of the system size. One possible explanation for this observation is that the human mobility, and not the epidemic vector (the mosquito), is the major mechanism responsible for the dissemination of the virus.
Human movement seems to be a crucial factor in the transmission dynamics, particularly in large tropical cities where successive dengue epidemics have been recorded, suggesting that the mosquito is highly adapted, with low intra-urban infestation differentials.
In addition, we also compared the disease transmission characteristics of dengue outbreaks in Fortaleza with the ones in Thailand. Our comparison showed that the number of dengue cases in Fortaleza is much higher compared to the outbreak data of Ref.~\cite{pandey2013comparing}. We also found that the obtained basic reproduction number of dengue outbreaks in Fortaleza is within the confidence interval of the value mentioned in Ref.~\cite{pandey2013comparing}. This means that the average number of secondary dengue cases originating from one infection is almost identical in both locations.
Moreover, we classified epidemic and non-epidemic years based on an analysis of spatio-temporal correlations and their corresponding distributions. Finally, we found that the spatial-temporal correlations are of the size of the system likely due to human mobility.
\renewcommand\thefigure{\thesection.\arabic{figure}}    
\setcounter{figure}{0}
\appendix
\section{Bayesian Markov chain Monte Carlo}
\label{sec:app_sir}
To compute the parameter distributions of the SIR model defined by Eqs.~\eqref{eq:sir_s}-\eqref{eq:sir_r} and \eqref{eq:sir_c}, we have to determine the probability distribution $P\left(\theta|D\right)$ of $\theta=\left(\beta,\gamma_H,p,r_0\right)$ given our data on the cumulative number of reported dengue cases $D$. According to Bayes' theorem, we express the posterior distribution
\begin{equation}
P\left(\theta|D\right)\propto P\left(D|\theta\right)P\left(\theta\right),
\label{eq:bayes}
\end{equation}
in terms of the likelihood function $P\left(D|\theta\right)$ and the prior parameter distribution $P\left(\theta\right)$. We assume that the likelihood function $P\left(D|\theta\right)$ is a Gaussian distribution of the error $E$ with zero mean and variance $\sigma^2$, i.e.,
\begin{equation}
P\left(D|\theta\right) \propto \exp\left(-\frac{E^2}{2 \sigma^2} \right).
\label{eq:likelihood}
\end{equation}
To compute the error, we discretize the infection period in monthly time intervals described by the times $t_i$ with $i\in\{1,2,\dots,12\}$ and evaluate $y_i(\theta)=C(t_i)/N$ for a given $\theta$. In the next step, we compare the prediction of our model $y_i(\theta)$ with the actual cumulative number of reported dengue cases $D_i$ at time $t_i$ and use the least square error
\begin{equation}
E^2=\sum_{i=1}^{12} \left[D_i - y_i\left(\theta\right)\right]^2
\label{eq:error}
\end{equation}
as our error estimate in Eq.~\eqref{eq:likelihood}~\cite{pandey2013comparing}. For a given prior parameter distribution $P\left(\theta \right)$, we compute the posterior distribution $P\left(D|\theta\right)$ using Bayesian
Markov chain Monte Carlo sampling with a Metropolis update scheme~\cite{pandey2013comparing,gelman2013bayesian}. After initializing the parameter vector with $\theta^0$ drawn from the prior distribution $P\left(\theta\right)$, the $n$th iteration of the algorithm is defined as follows:
\begin{enumerate}
\item A new parameter set $\theta^\ast$ is drawn from the proposal distribution $J\left(\theta^\ast | \theta^{n} \right)$.
\item The acceptance probability for $\theta^\ast$ is computed according to (Metropolis algorithm)
\begin{equation}
r=\min\left(\frac{P\left(\theta^\ast|D\right)}{P\left(\theta^n|D\right)},1\right)=\min\left(\frac{P\left(D|\theta^\ast\right)P\left(\theta^\ast\right)}{P\left(D|\theta^n\right)P\left(\theta^n\right)},1\right).
\label{eq:metropolis_app}
\end{equation}
\item Draw a random number $\epsilon\sim U(0,1)$ and set
\begin{equation}
\theta^{n+1} = \begin{cases} \theta^\ast &\text{if } \epsilon < r, \\
\theta^{n} & \text{otherwise.} \end{cases} 
\end{equation}
\end{enumerate}
This update procedure implies that a new parameter set is always accepted if the new likelihood function value is greater or equal than the one of the previous iteration, i.e., if $P\left(D|\theta^\ast\right)\geq P\left(D|\theta^{n}\right)$. For the described Metropolis algorithm, the proposal distribution $J\left(\theta^\ast | \theta^{n} \right)$ must be symmetric. According to Ref.~\cite{gelman2013bayesian}, a multivariate Gaussian distribution
\begin{equation}
J\left(\theta^\ast | \theta^{n} \right) \sim N\left(\theta^{n} | \lambda^2 \Sigma \right)
\end{equation}
may be used as proposal distribution. The covariance matrix is denoted by $\Sigma$ and the corresponding scaling factor by $\lambda$. Every 500 iterations, the covariance matrix and the scaling factor are updated using the following update procedure~\cite{pandey2013comparing,gelman2013bayesian}:
\begin{align}
\begin{split}
\Sigma_{k+1} &= p \Sigma_{k}+(1-p) \Sigma ^\ast, \\
\lambda_{k+1} &= \lambda_{k}\exp\left(\frac{\alpha^\ast-\hat{\alpha}}{k}\right),
\end{split}
\end{align}
where $\Sigma^\ast$ and $\alpha^\ast$ are the covariance matrix and the acceptance rate of the last 500 iterations, respectively. Furthermore, $p=0.25$ and the initially $\Sigma_0=\mathbb{I}$ and $\lambda_0=2.4/\sqrt{d}$ where $\mathbb{I}$ is the $d\times d$ identity matrix and $d$ the number of estimated parameters. The target acceptance rate is~\cite{gelman2013bayesian}
\begin{equation}
\hat{\alpha}= \begin{cases}0.44 &\text{if }d=1, \\
0.23& \text{otherwise.} \end{cases} 
\end{equation}
\begin{figure}
\begin{minipage}{0.49\textwidth}
\centering
\includegraphics[width=\textwidth]{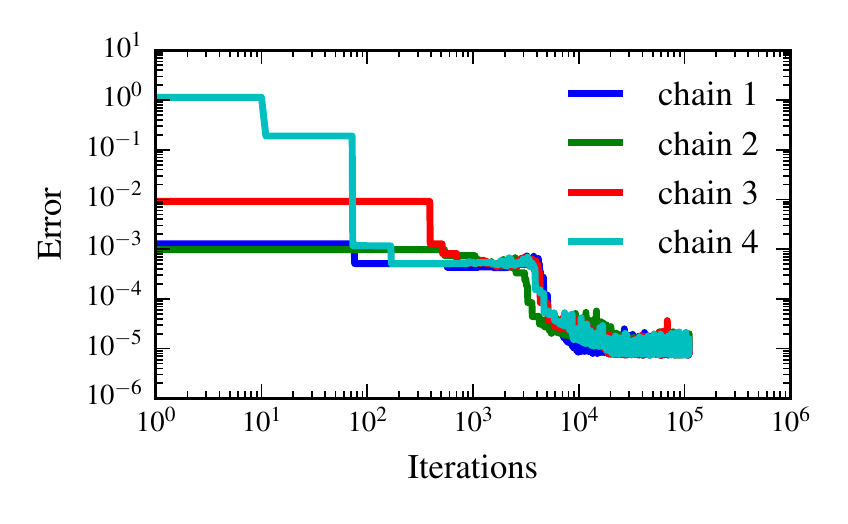}
\end{minipage}
\begin{minipage}{0.49\textwidth}
\centering
\includegraphics[width=\textwidth]{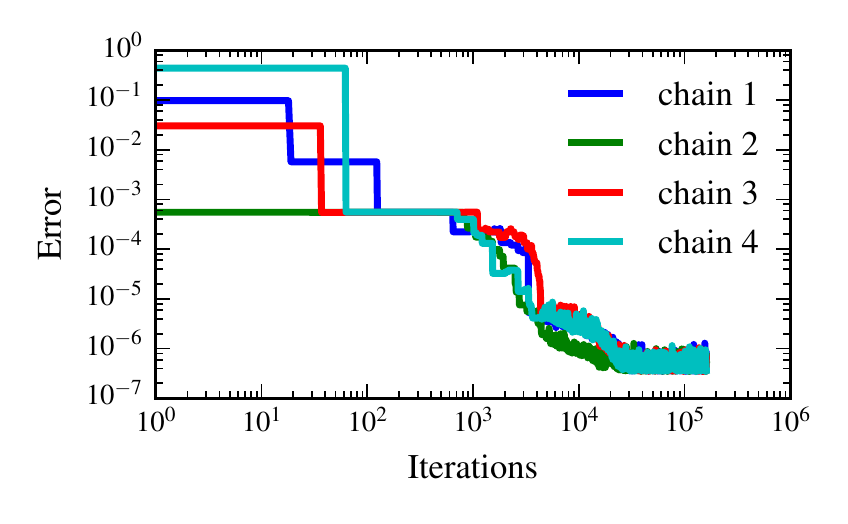}
\end{minipage}
\caption{\textbf{Convergence of Bayesian
Markov chain Monte Carlo parameter estimation.} For both epidemic years 2012 (left) and 2015 (right), four different chains with different initial conditions have been considered to perform Bayesian
Markov chain Monte Carlo parameter estimation. The algorithm is considered to be converged to the posterior distribution if the variance between the chains is similar to the variance within the chains (Gelman-Rubin test)~\cite{gelman2013bayesian}.} 
\label{fig:sir_convergence}
\end{figure}

For evaluating Eq.~\eqref{eq:metropolis_app}, we have to compute the least-square error for the new proposed parameter set $\theta^\ast$ according to Eq.~\eqref{eq:error} to then obtain the likelihood function value $P\left(D|\theta^\ast\right)$ based on Eq.~\eqref{eq:likelihood}. To obtain good convergence behavior, we set the variance of our likelihood distribution to $\sigma^2=1/20$. Convergence is measured in terms of the Gelman-Rubin test~\cite{gelman2013bayesian}. Therefore, we consider four independent Markov chains. Every chain is initialized with a different random parameter set which is drawn from the corresponding uniform distributions. The posterior parameter distribution is considered to be converged if the variance between the chains is similar to the variance within the chain (Gelman-Rubin test).
In Fig.~\ref{fig:sir_convergence}, we illustrate the convergence behavior. After reaching convergence, we produce $10^5$ more samples without updating the covariance matrix and construct the posterior parameter distribution based on every $5$th sample. 

For both epidemic years, we use four independent Markov chains with different initial parameter sets $\theta=\left(\beta, \gamma_H, p, r_0\right)$ that are drawn from uniform posterior distributions. If the variance between the chains is similar to the variance within the chain (Gelman-Rubin test), we consider the posterior distribution to be converged. Every new proposed set of parameters $\theta^\ast$ is accepted with probability (Metropolis algorithm)~\cite{gelman2013bayesian}
\begin{equation}
r=\min\left(\frac{P\left(\theta^\ast|D\right)}{P\left(\theta|D\right)},1\right)=\min\left(\frac{P\left(D|\theta^\ast\right)P\left(\theta^\ast\right)}{P\left(D|\theta\right)P\left(\theta\right)},1\right).
\label{eq:metropolis}
\end{equation}
The resulting posterior distributions are shown in Fig.~\ref{fig:sir_posterior}. We find similar posterior distributions of the parameters $\beta$, $\gamma_H$ and $r_0$ for both epidemic years. The distributions of $p$ (fraction of infected individuals that were diagnosed with dengue and reported to the health officials) are different due to the larger number of reported dengue cases in 2012 (38197) compared to 2015 (26176).\\\\

%
%
%
%\section*{Declarations}
\noindent{\bf\large Declarations}
%
%
%
%\subsection*{Availability of data and material}
\\\\\noindent{\bf Availability of data and material}\\\\
All data generated or analysed during this study are included in this published article [and its supplementary information files].
%
%
%
%\subsection*{Competing interests}
\\\\\noindent{\bf Competing interests}\\\\
The authors declare no competing interest.
%
%
%
%\subsection*{Funding}
\\\\\noindent{\bf Acknowledgements}\\\\
We acknowledge financial support from the Brazilian agencies CNPq, CAPES, the FUNCAP Cientista Chefe program, and from the National Institute of Science and Technology for Complex Systems (INCT-SC) in Brazil. LB acknowledges financial support from the SNF Early Postdoc. Mobility fellowship on ``Multispecies interacting stochastic systems in biology'' and the Army Research Office (W911NF-18-1-0345).
%
%
%
%\subsection*{Authors' contributions}
\\\\\noindent{\bf Authors' contributions}\\\\
All authors designed research. GSS and ASLN collected and prepared the Dengue data sets. SDRS, LB, and JPdCN processed the Dengue data. All authors analyzed and interpreted the Dengue data. SDSR, LB, ASLN, HH and JSA wrote the manuscript. All authors read and approved the final manuscript.
\bibliography{refs}
\bibliographystyle{vancouver}
\end{document}